\documentclass[10pt]{article}

\usepackage{amsmath}
\usepackage{amssymb}
\usepackage{mathrsfs}
\usepackage{graphics}

\usepackage{duotenzor}

\newcommand{\negs }{\hspace{-1pt}}

\begin{document}

\title{\textbf{Reconstructing Quantum Theory}}
\author{Lucien Hardy\\
\textit{Perimeter Institute,}\\
\textit{31 Caroline Street North,}\\
\textit{Waterloo, Ontario N2L 2Y5, Canada}}
\date{}

\maketitle

\begin{abstract}
I provide operational postulates for quantum theory.  These involve certain operational notions.  Systems come in different types, $\mathsf{a, b, c, \dots}$.  A \emph{maximal set of distinguishable states} is any set containing the maximum number, $N_\mathsf{a}$, of states for which there exists some measurement, called a \emph{maximal measurement}, which can identify which state from the set we have in a single shot.  A \emph{maximal effect} is associated with each result of a maximal measurement.  An \emph{informational face} is the full set of states that give rise only to some subset of outcomes of some maximal measurements (it corresponds to constraining the system to have some reduced information carrying capacity). States are represented by vectors whose $K_\mathsf{a}$ entries are probabilities. A set of states is said to be \emph{non-flat} if it is a spanning subset of some informational face. A \emph{filter} is a transformation that passes unchanged those states in a given informational face while blocking those states in the complement informational face (that would give rise only to outcomes in the complement outcome set of the maximal measurement).
Classical probability theory and quantum theory are the only two theories consistent with the following set of postulates.
\begin{description}
\item[P1] \emph{Logical sharpness.} There is a one-to-one map between pure states and maximal effects such that we get unit probability.
\item[P2] \emph{Information locality.} A maximal measurement is effected on a composite system if we perform maximal measurements on each of the components (or, equivalently, $N_\mathsf{ab}=N_\mathsf{a}N_\mathsf{b}$). 
\item[P3] \emph{Tomographic locality.} The state of a composite system can be determined from the statistics collected by making measurements on the components (or, equivalently, $K_\mathsf{ab}=K_\mathsf{a}K_\mathsf{b}$).
\item[P4$'$] \emph{Permutability.} There exists a reversible transformation on any system effecting any given permutation of any given maximal set of distinguishable states for that system.
\item[P5] \emph{Sturdiness.} Filters are non-flattening.
\end{description}
To single out quantum theory we need only add any requirement that is inconsistent with classical probability theory and consistent with quantum theory.
%
\end{abstract}

\section{Motivation}

The standard axioms of QT are rather ad hoc.  Where does this structure come from?  Can we write down natural axioms, principles, laws, or postulates from which can derive this structure?
Compare with the Lorentz transformations and Einstein's two postulates for special relativity.  Or compare with Kepler's Laws and Newton's Laws.  The standard axioms of quantum theory look rather ad hoc like the Lorentz transformations or Kepler's laws.  Can we find a natural set of postulates for quantum theory that are akin to Einstein's or Newton's laws?

The real motivation for finding deeper postulates for quantum theory is that it may help us go beyond quantum theory to a theory of quantum gravity (just as Einstein's work helped him go beyond special relativity to his theory of General Relativity).  It is in the finding of new physics that we can expect a real payoff of this program.

In \cite{hardy2011reformulating} I showed how classical probability theory and quantum theory are the only two theories consistent with the  set of postulates given above in the abstract. In this chapter I will explain the meaning of these postulates and indicate how the main steps of the proof work.  The reconstruction takes place in the context of the circuit framework which I will describe.

\section{A personal history of reconstruction}

A dozen years or so years ago Christopher Fuchs implored the community to \lq\lq find an information theoretic reason" for axioms of QT (in multiple talks and a few papers \cite{fuchs2001quantum, fuchs2002quantum}).  Further, Chris Fuchs and Gilles Brassard invited me to a workshop in Montreal in 2000 on this issue amongst others (see notes in \cite{fuchs2009coming}).  I accepted their invitation but was, in the event, unable to attend. However, I was already hooked.  The work I began preparing for that workshop led to my paper \lq\lq Quantum theory from five reasonable axioms" \cite{hardy2001quantum}.   In modern form (see \cite{hardy2009foliable}), the axioms given there can be stated in the following way:
\begin{description}
\item[Information] Systems having, or constrained to have, a given information carrying
capacity have the same properties.
\item[Information locality] Same as {\bf P2} above.
\item[Tomographic locality] Same as {\bf P3} above.
\item[Continuous reversibility] There exists a continuous reversible transformation between any
pair of pure states.
\item[Simplicity] States are specified by the smallest number of probabilities consistent
with the other axioms.
\end{description}
The simplicity axiom stands out as being less reasonable than the others.  If we drop it then we may get a hierarchy of theories.  This leads to two possibilities.  Either there do exist higher theories in this hierarchy or there do not.  For many years I tried to find such theories, and I tried to prove that such theories do not exist.  I also tried to find other reasonable axioms that rule out higher theories in this hierarchy.  It was not until 2009 that progress was made by others.   In 2009 Chiribella, D'Ariano, and Perinotti (CDP) \cite{chiribella2010probabilistic} showed how considerations concerning teleportation can be used to get rid of the need for a simplicity axiom in certain contexts.  In 2010 \cite{chiribella2010informational} they found a set of postulates for quantum theory which, by virtue of the techniques developed in \cite{chiribella2010probabilistic}, did not require a simplicity axiom.  Independently Daki\'c and Brukner \cite{dakic2009quantum} showed how one can replace the simplicity axiom with the assumption that any state for a two level system can be written as a mixture of perfectly distinguishable states (modulo some technical problems in their proof arising from the unfortunate existence of a subgroup of SO(7) that is transitive on the 6-sphere).  In 2010 Masanes and M\"uller \cite{masanes2010derivation} showed how to replace the simplicity axiom with a different axiom saying that all mathematically well defined measurements for a two-level system are allowed.   The axiom sets of Daki\'c and Brukner, and of Masanes and M\"uller are slight modifications on my original axiom set from 2001.  The axioms of CDP are quite different (except for the assumption of tomographic locality). Masanes, M\"uller, Augusiak, and P\'erez-Garc\'a provide another set of axioms employing tomographic locality, continuous reversibility, and another axiom concerning the existence of an informational unit.  Another set of axioms using tomographic locality was given by Marco Zaopo \cite{zaopo2012information}.

There has been much work recently by many people along less related lines (Fuchs \cite{fuchs2010quantum}, Goyal \cite{goyal2008information}, Wilce \cite{wilce2009four}, Rau \cite{rau2009quantum, rau2010measurement}, Fivel \cite{fivel2010derivation}, \dots).  Further, there is, in fact, a long history of attempts to reconstruct QT (von Neumann \cite{von1996mathematical}, Mackey \cite{mackey1963mathematical}, Birkoff and von Neumann \cite{birkhoff1936logic}, Zierler \cite{zierler1975axioms}, Piron \cite{piron1964axiomatique} Ludwig \cite{ludwig1985axiomatic}, Rovelli \cite{rovelli1996relational}, and many others).

Many of these reconstruction attempts employ the so called \lq\lq convex probabilities framework".  This goes back to originally to Mackey and has been worked on (and sometimes rediscovered) by many others since including Ludwig \cite{ludwig1985axiomatic}, Davies and Lewis \cite{davies1970operational}, Gunson \cite{gunson1967algebraic}, Mielnik \cite{mielnik1969theory}, Araki \cite{araki1980characterization}, Gudder {\it et al.\ } \cite{gudder1999convex}, Foulis and Randall \cite{foulis1979empirical}, Fivel \cite{fivel1994interference} as well as more recent incarnations \cite{hardy2001quantum, barrett2007information}.

The circuit framework used here \cite{hardy2010formalism, hardy2011reformulating} (see also \cite{hardy2009foliable, hardy2009operational}) might be regarded as a marriage of the convex probabilities framework and the pictorial (or categorical) approach of Abramsky and Coecke \cite{abramsky2004categorical, coecke2010quantum}.   A similar framework has been developed by Chiribella, D'Ariano, and Perinotti \cite{chiribella2010probabilistic}.   The pictorial approach of Abramsky and Coecke is important because of its emphasis on composition as a basic primitive.

In 2002 Clifton, Bub, and Halvorson \cite{clifton2003characterizing} were inspired by a suggestion of Fuchs and Brassard to take a different approach to reconstructing quantum theory. They showed that some features of quantum theory follow if one imposes no-bit-commitment, no-broadcasting, and no-signalling within the $C^*$ algebraic framework (rather than the convex probabilities framework).

\section{The circuit framework}

In this section we will present the circuit framework.  The basic idea is that circuits can be built from operations. An operation corresponds to one use of an apparatus with some particular outcome, or subset of possible outcomes specified.  Operations have some number of systems as inputs and some number of systems as outputs.  We can wire together operations.  If we have no open inputs or outputs left over then we have a circuit.  We employ three background assumptions for this framework. The main one is that we can associate a probability with a circuit (the joint probability that the outcomes are in the associated outcome sets on each operation).

\subsection{Operations}

We can notate an operation diagrammatically or symbolically as follows.
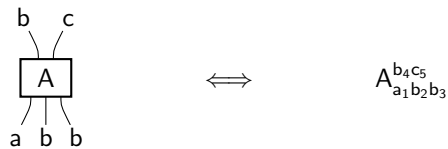
\begin{equation}\label{operationA}
\begin{Diagram}{0}{-0.80}
\Opbox{A}{3,3}
\inwire[-5]{A}{1}\Opsymbol{a}
\inwire{A}{2}\Opsymbol{b}
\inwire[5]{A}{3}\Opsymbol{b}
\outwire[-5]{A}{1.5}\Opsymbol{b}
\outwire[5]{A}{2.5}\Opsymbol{c}
\end{Diagram}
~~~~~~~~~~~~\Longleftrightarrow ~~~~~~~~~~~~~\mathsf{A_{a_1b_2b_3}^{b_4c_5}}
\nonumber\end{equation}
The integer subscripts in the symbolic representation will be used to show where the wires go and have no significance beyond this.
An operation, $\mathsf{A}$, corresponds to one use of an apparatus and has the following features.
\begin{itemize}
\item {\it Inputs and outputs.}  Come in various types, $\mathsf a$, $\mathsf b$, \dots. The inputs correspond to wires going in the bottom of the box in the diagrammatic notation, or subscripts in the symbolic notation.  The outputs correspond to wires coming out the top and to superscripts.
\item {\it A setting}, $\mathsf{s(A)}$.  We can think of this as corresponding to certain positions for knobs, buttons, and dials that may be on the apparatus.
\item {\it An outcome set}, $\mathsf{o(A)}$.  This is a subset of all the possible outcomes for this use of the apparatus.
\end{itemize}
If $\mathsf{x_A\in o(A)}$ then we say operation $\mathsf{A}$ \lq\lq happened".  If we have a different setting, or a different outcome set, then we have a different operation and should notate this with a different letter (e.g.\ $\mathsf B$ rather than $\mathsf A$).

\subsection{Wires}

Outputs can be connected to inputs by wires.
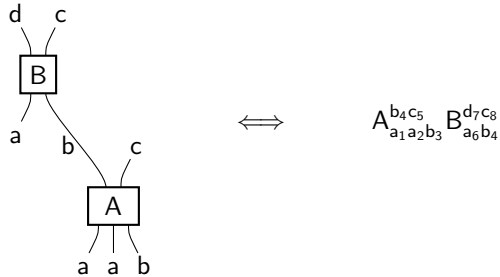
\begin{equation}\label{ABwired}
\begin{Diagram}{0}{-1.2}
\Opbox{A}{7,0}
\Opbox[2]{B}{3,7}
\inwire[-5]{A}{1}\Opsymbol{a}
\inwire{A}{2}\Opsymbol{a}
\inwire[5]{A}{3}\Opsymbol{b}
\wire{A}{B}{1.5}{2}\opsymbol{b}
\outwire[5]{A}{2.5}\Opsymbol{c}
\inwire[-5]{B}{1}\Opsymbol{a}
\outwire[-5]{B}{1}\Opsymbol{d}
\outwire[5]{B}{2}\Opsymbol{c}
\end{Diagram}
~~~~~~~~ \Longleftrightarrow ~~~~~~~~~ \mathsf{A_{a_1a_2b_3}^{b_4c_5}}\mathsf{B_{a_6b_4}^{d_7c_8}}
\nonumber\end{equation}
Note how the wire linking the two boxes corresponds to the repeated integer (the 4 on $\mathsf{b}_4$).  These diagrams are interpreted graphically.  In particular, vertical position has no meaning.  We can distort the graph in any way we wish without changing the physical meaning so long as the wires remain attached to the same positions on the boxes and the boxes maintain their orientations.

There are certain wiring rules.
\begin{itemize}
\item {\it One wire:} At most one wire can be connected to any given input or output.
\item{\it Type matching:} Wires can connect inputs and outputs of the same type.
\item{\it No closed loops:} If we trace from output to input along wires through the operations then we cannot get back to the operation we started at.
\end{itemize}
The last rule is to rule out closed time like loops.

\subsection{Fragments}

The most general object we can consider is a collection of operations wired together.  For example,
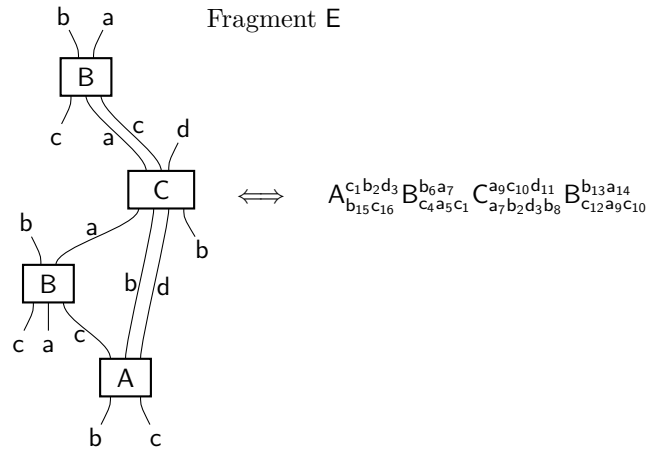
\begin{equation}\label{fragmentE}
\begin{Diagram}[1]{-4}{-1}
\Opbox{A}{4.1,-6}\Opbox{B}{0,-1}\Opbox[4]{C}{6,4}\opbox{F}{2,10}\opsymbol{B}
\inwire[-5]{A}{1} \Opsymbol{b} \inwire[5]{A}{3} \Opsymbol{c}
\inwire[-5]{B}{1}\Opsymbol{c}\inwire{B}{2}\Opsymbol{a}\wire{A}{B}{1}{3}\opsymbol{c}\wire{A}{C}{2}{2}\opsymbol{b}\wire{A}{C}{3}{3}\otherside\opsymbol{d}
\outwire[-5]{B}{1.5}\Opsymbol{b}\wire{B}{C}{2.5}{1}\opsymbol{a}
\inwire[6]{C}{4}\Opsymbol{b}
\wire{C}{F}{1.5}{2}\opsymbol{a}\wire{C}{F}{2.5}{3}\otherside\opsymbol{c}\outwire[5]{C}{3}\Opsymbol{d}
\inwire[-5]{F}{1}\Opsymbol{c}
\outwire[-5]{F}{1.5}\Opsymbol{b}\outwire[5]{F}{2.5}\Opsymbol{a}
\placelatex{12,13}{\text{Fragment} \ensuremath{\mathsf E}}
\end{Diagram} \hspace{-1.6cm}
\Longleftrightarrow ~~~~ \mathsf{ A^{c_1b_2d_3}_{b_{15}c_{16}} B_{c_4a_5c_1}^{b_6a_7} C_{a_7b_2d_3b_8}^{a_9c_{10}d_{11}} B_{c_{12} a_9 c_{10}}^{b_{13}a_{14}}}
\nonumber\end{equation}
Such objects are called fragments (as they are fragments of circuits).  In general fragments may have open inputs and outputs.  Fragments have
\begin{itemize}
\item {\it A setting}, $\mathsf {s(E)}$, given by specifying the setting on each operation.
\item {\it An outcome set}, $\mathsf {o(E)}$, (equals $\mathsf{o(A)\times o(B) \times o(C) \times o(A)}$ in this case). We say the fragment \lq\lq happened" if the outcome is in the outcome set.
\item {\it A wiring}, $\mathsf {w(E)}$, given  by specifying the input/output pairs which are wired together.  
\end{itemize}

\subsection{Circuits}

Circuits have no open inputs or outputs.  For example,
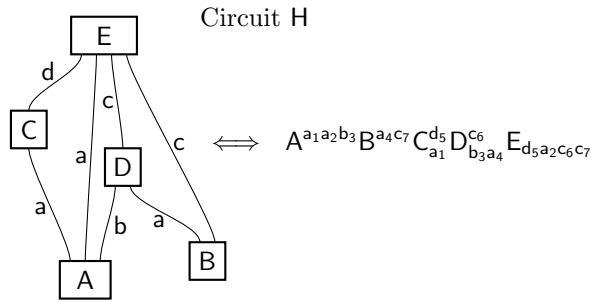
\begin{equation}
\begin{Diagram}{0}{-1.4}
\Opbox{A}{8,-2}\Opbox[2]{B}{14.5,-1}\Opbox[2]{C}{5,6}\Opbox[2]{D}{10,4}\Opbox[4]{E}{9,11}
\wire{A}{C}{1}{1.5}\opsymbol{a}\wire{A}{E}{2}{2}\opsymbol{a}\wire{A}{D}{3}{1}\otherside\opsymbol[5,0]{b}
\wire{B}{D}{1}{2}\opsymbol{a}\wire{B}{E}{2}{4}\otherside\opsymbol{c}\wire{D}{E}{1.5}{3}\opsymbol{c}\wire{C}{E}{1.5}{1}\opsymbol[0,8]{d}
\placelatex{17,12}{\text{Circuit} \ensuremath{\mathsf H} }
\end{Diagram}
\hspace{-1.5cm}
\Longleftrightarrow ~~  \mathsf{   A^{a_1a_2b_3} B^{a_4c_7} C_{a_1}^{d_5} D_{b_3a_4}^{c_6} E_{d_5a_2c_6c_7}  }
\nonumber\end{equation}
Circuits are special cases of fragments.  Circuits have
\begin{itemize}
\item {\it A setting}, $\mathsf {s(H)}$, given by specifying the setting on each operation.
\item {\it An outcome set}, $\mathsf {o(H)}$, given by specifying the outcome set at each operation (equals $\mathsf{o(A)\times o(B) \times o(C) \times o(D) \times o{E}}$ in this case). We say the fragment \lq\lq happened" if the outcome is in the outcome set.
\item {\it A wiring}, $\mathsf {w(E)}$, given by specifying which input/output pairs are wired together.  
\end{itemize}

\subsection{Preparations, results, and transformations}

A {\bf preparation} is a fragment having only outputs. Here are some examples:
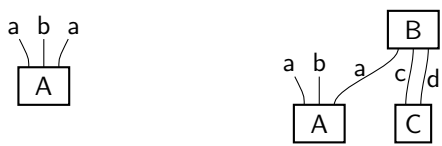
\[
\begin{Diagram}{0}{0.5}
\Opbox{A}{0,0}
\outwire[-5]{A}{1} \Opsymbol{a} \outwire{A}{2} \Opsymbol{b} \outwire[5]{A}{3} \Opsymbol{a}
\end{Diagram}
~~~~~~~~~~~~~~~~~~~~~
\begin{Diagram}{0}{0}
\Opbox{A}{0,0} \Opbox{B}{5,5} \Opbox[2]{C}{5,0}
\wire{A}{B}{3}{1} \opsymbol{a}   \wire{C}{B}{1}{2} \opsymbol{c} \wire{C}{B}{2}{3}\otherside\opsymbol{d}
\outwire[-5]{A}{1} \Opsymbol{a} \outwire{A}{2} \Opsymbol{b}
\end{Diagram}
\]
We will associate \emph{states} with preparations.

A {\bf result} is a fragment having only inputs.  Here are some examples:
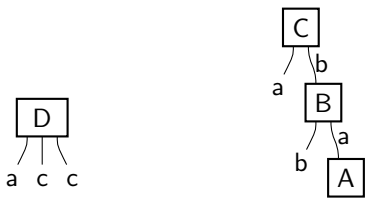
\[
\begin{Diagram}{0}{0.8}
\Opbox{D}{0,0}
\inwire[-5]{A}{1} \Opsymbol{a} \inwire{A}{2} \Opsymbol{c} \inwire[5]{A}{3} \Opsymbol{c}
\end{Diagram}
~~~~~~~~~~~~~~~~~~~~
\begin{Diagram}{0}{0}
\Opbox[2]{A}{0,0} \Opbox[2]{B}{-1.2,4} \Opbox[2]{C}{-2.4,8}
\wire{A}{B}{1}{2}\otherside \opsymbol{a} \wire{B}{C}{1}{2} \otherside\opsymbol{b}
\inwire[-5]{B}{1} \Opsymbol{b} \inwire[-5]{C}{1} \Opsymbol{a}
\end{Diagram}
\]
we will associate \emph{effects} with results.   A {\bf measurement} is a collection of results corresponding to the same setup with disjoint outcome sets whose union is the set of all outcomes for this setup.

A {\bf transformation} is a fragment that has inputs and outputs that is \emph{used in transformation mode}.
Here are some examples
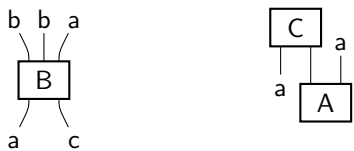
\[
\begin{Diagram}{0}{0}
\Opbox{B}{0,0}
\inwire[-5]{B}{1}\Opsymbol{a}\inwire[5]{B}{3} \Opsymbol{c}
\outwire[-5]{B}{1}\Opsymbol{b}\outwire{B}{2}\Opsymbol{b}\outwire[5]{B}{3}\Opsymbol{a}
\end{Diagram}
~~~~~~~~~~~~~~~~~~~~
\begin{Diagram}{0}{-0.3}
\Opbox{A}{0,0} \Opbox{C}{-1.6,4} \wire{A}{C}{1}{3}
\inwire{C}{1} \Opsymbol{a} \outwire{A}{3} \Opsymbol{a}
\end{Diagram}
\]
A fragment is used in transformation mode if we do not feed outputs into inputs on this fragment (either directly or indirectly).

A transformation, $\mathsf{B_{a_1}^{a_2}}$, is reversible if there exists another transformation, $\mathsf{\tilde B_{a_2}^{a_3}}$, such that $\mathsf{B_{a_1}^{a_2}\tilde B_{a_2}^{a_3}}$ is the identity transformation.  Note that the identity transformation is defined to have the property that, if it inserted on any wire in any circuit, then the probability for that circuit remains unchanged.

\subsection{The first background assumption}

We need three background assumptions in setting up the circuit framework.  The first is the following.
\begin{quote}
{\bf Assump 1.} We can associate a probability with any given circuit (the probability that the circuit \lq\lq happens"), and this probability depends
only on the specification of the given circuit (the knob settings and outcome sets at the operations, and the wiring).
\end{quote}
For example,
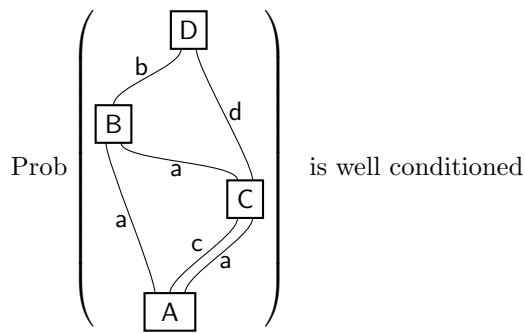
\begin{equation}
\text{Prob}\left(
\begin{Diagram}{0}{-2}
\Opbox{A}{0,0} \Opbox[2]{C}{4,6} \Opbox[2]{B}{-3,10} \Opbox[2]{D}{1,15}
\wire{A}{B}{1}{1}\opsymbol{a} \wire{A}{C}{2}{1}\opsymbol[0,6]{c} \wire{A}{C}{3}{2}\otherside\opsymbol{a} \wire{C}{B}{1}{2}\opsymbol{a}
\wire{C}{D}{2}{2}\otherside\opsymbol[4,0]{d} \wire{B}{D}{1.5}{1}\opsymbol[0,6]{b}
\end{Diagram}
\right)
~~ \text{is well conditioned}
\nonumber\end{equation}
Note that we make this assumption for circuits, not for fragments generally. Indeed, a fragment with open inputs and/or outputs cannot be expected to satisfy this since the probability may depend on what is done with these open ports.

\subsection{The state}

Want to associate a state with a preparation
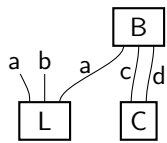
\[
\begin{Diagram}{0}{0}
\opbox{A}{0,0}\opsymbol{L} \Opbox{B}{5,5} \Opbox[2]{C}{5,0}
\wire{A}{B}{3}{1} \opsymbol{a}   \wire{C}{B}{1}{2} \opsymbol{c} \wire{C}{B}{2}{3}\otherside\opsymbol{d}
\outwire[-5]{A}{1} \Opsymbol{a} \outwire{A}{2} \Opsymbol{b}
\end{Diagram}
\]
There exist many results which complete this into a circuit.  Here are a few examples:
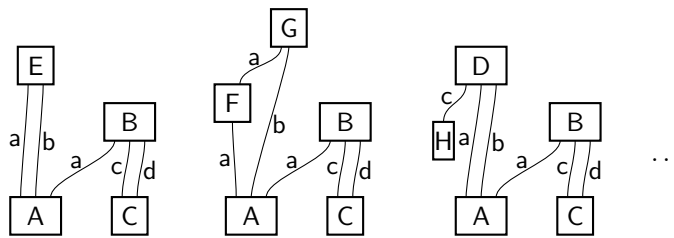
\[
\begin{Diagram}{0}{-0.9}
\Opbox{A}{0,0} \Opbox{B}{5,5} \Opbox[2]{C}{5,0}
\wire{A}{B}{3}{1} \opsymbol{a}   \wire{C}{B}{1}{2} \opsymbol{c} \wire{C}{B}{2}{3}\otherside\opsymbol{d}
\Opbox[2]{E}{0,8} \wire{A}{E}{1}{1}\opsymbol{a} \wire{A}{E}{2}{2} \otherside \opsymbol{b}
\end{Diagram}
~~~~~
\begin{Diagram}{0}{-0.9}
\Opbox{A}{0,0}\Opbox{B}{5,5} \Opbox[2]{C}{5,0}
\wire{A}{B}{3}{1} \opsymbol{a}   \wire{C}{B}{1}{2} \opsymbol{c} \wire{C}{B}{2}{3}\otherside\opsymbol{d}
\Opbox[2]{F}{-1,6}\Opbox[2]{G}{2,10}\wire{F}{G}{2}{1} \opsymbol{a}
\wire{A}{F}{1}{1.5}\opsymbol{a} \wire{A}{G}{2}{1.5} \otherside \opsymbol{b}
\end{Diagram}
~~~~~
\begin{Diagram}{0}{-0.9}
\Opbox{A}{0,0} \Opbox{B}{5,5} \Opbox[2]{C}{5,0}
\wire{A}{B}{3}{1} \opsymbol{a}   \wire{C}{B}{1}{2} \opsymbol{c} \wire{C}{B}{2}{3}\otherside\opsymbol{d}
\Opbox{D}{0,8}  \Opbox[1]{H}{-2,4} \wire{H}{D}{1}{1} \opsymbol{c}
\wire{A}{D}{1}{2}\opsymbol{a} \wire{A}{D}{2}{3} \otherside \opsymbol{b}
\end{Diagram}
~~~~ \dots
\]
The state associated with a preparation should enable us to predict the probability for every circuit containing this preparation.

We could specify the state associated with a preparation by giving a list of probabilities for every circuit made with this preparation.   This would be a very long list and rather cumbersome to work with. However, physical theories typically relate different quantities.  Consequently it should be possible to pick out a subset results such that specifying the probabilities for just the circuits containing the given preparation and results from this subset is sufficient to allow us to calculate the probability for any other circuit containing this preparation.  For example, in Quantum Theory we can calculate all the probabilities for a spin-half particle from just the probabilities
\[
\left( \begin{array}{c} p_{x+} \\  p_{y+} \\ p_{z+} \\ p_{z-} \end{array} \right)
\]
as these suffice to determine the density matrix for this system.  In fact, these probabilities suffice to determine the elements of the density matrix by linear equations.  We will insist on linearity in what follows.  This is well justified when one considers taking mixtures of states (see Appendix B of \cite{hardy2011reformulating} for example).  We call the choice of results used to specify the state {\bf fiducial results}.  In general this choice is not unique.

\subsection{Using Fiducial results to define states}

It is worth paying attention to the font used in the notation below.  Consider preparations of the form $\mathsf{A^{a_1}}$.  We choose a fiducial set of results
\[
\mathsf{X}_\mathsf{a_1}^{a_1} ~~~\text{for} ~a_1 = 1 ~\text{to}~ K_\mathsf{a}
\]
The state associated with preparation $\mathsf{A^{a_1}}$ is given by
\[
A^{a_1} := \text{Prob}(\mathsf{A^{a_1}} \mathsf{X}_\mathsf{a_1}^{a_1} )
\]
A fiducial set is a minimal set such that, for any result, $\mathsf{B_{a_1}}$ there exists an {\bf effect} $B_{a_1}$ such that
\[ \text{Prob}(\mathsf{A^{a_1}B_{a_1}}) = A^{a_1} B_{a_1} ~~~~ (\text{summation over}~a_1~\text{implied})\]
This is linear.  If we allow arbitrary mixtures then must have linearity here \cite{hardy2011reformulating}.  However, even if we do not allow arbitrary mixtures, we are free to consider only linear relations of this type even though there may be a more efficient non-linear expression.  Associated with {\bf preparation} $\mathsf{A^{a_1}}$ is the {\bf state}, $A^{a_1}$.  This is a list of $K_\mathsf{a}$ fiducial probabilities from which all other probabilities can be calculated.  Associated with the {\bf result} $\mathsf{B_{a_1}}$ is the {\bf effect} $B_{a_1}$. This is a list of $K_\mathsf{a}$ real coefficients (which can be negative).

\subsection{Pure states}

A {\bf mixed state} is one that can be simulated by a mixture of preparations.  I.e.
\[ A^{a_1}= \lambda B^{a_1} + (1-\lambda) C^{a_1} \]
where $0<\lambda <1$ and $B^{a_1}\not= C^{a_1}$.   A {\bf pure state} is one that cannot be simulated by a mixture of preparations.  A transformation is {\bf non-mixing} if it preserves purity (up to normalization).

\subsection{Maximal sets}

A very important notion is that of a maximal set.
\begin{quote}
A {\bf maximal set of distinguishable states} is any set containing the maximum number, $N_\mathsf{a}$, of states for which there exists some measurement, called a {\bf maximal measurement}, which can identify which state from the set we have in a single shot.
\end{quote}
We also need the following notion.  
\begin{quote}
A {\bf maximal effect} is associated with each outcome of a maximal measurement.
\end{quote}
We can notate these notions diagrammatically as
\[
\text{Prob}\left(
\begin{Diagram}{0}{-0.8}
\opbox[3.4]{An}{0,0} \opsymbol{A\mathnormal{[n]}}
\opbox[3.4]{Bm}{0,6} \opsymbol{B\mathnormal{[\! m \!]}}
\wire{An}{Bm}{2.2}{2.2} \opsymbol{a}
\end{Diagram}
\right)
= \delta_{nm}
\]
or symbolically as
\[ A^{a_1}[n] B_{a_1}[m] = \delta_{nm}  \]
where $m,n=1$ to $N_\mathsf{a}$.

In quantum theory maximal sets of distinguishable states are associated with an orthonormal basis.  Then $N_\mathsf{a}$ is the dimension of the Hilbert space, maximal measurements correspond to non-degenerate observables, and maximal effects correspond to rank-one projectors.

In classical probability theory there is a unique maximal set of distinguishable states and it is usually understood to correspond to the underlying states of reality.

\subsection{Two more assumptions for framework}

We need two more background assumptions for the circuit framework.
\begin{quote}
{\bf Assump 2.}  There exists at least one type of system having $N_\mathsf{a}>1$ and $K_\mathsf{a}$ finite.
\end{quote}
Recall that $N_\mathsf{a}$ is the maximum number of states in a distinguishable set and $K_\mathsf{a}$ is the number of probabilities that must be provided to specify the state.
In quantum theory we note that systems having finite $N_\mathsf{a}$ also have finite $K_\mathsf{a}$.

We will give the third assumption without defining all the terms.
\begin{quote}
{\bf Assump 3.} If, for any accuracy $\delta>0$, there exists a fragment $A[\delta]$ that is operationally indiscernible from a given  hypothetical fragment, $\mathsf Q$, then there actually exists a fragment with the probabilistic properties of $\mathsf Q$.
\end{quote}
This assumption is used to obtain the property that the space of fragments is compact in an appropriate sense. The reader is referred to \cite{hardy2011reformulating} for more details.

\subsection{Permutation transformations}

We can define permutation transformations with respect to a given maximal set of distinguishable states
\[
\begin{Diagram}{0}{-1}
\opbox[5.2]{An}{0,0} \opsymbol{A\mathnormal{[n]}}
\opbox[3.4]{P}{0,5}   \opsymbol{P_\mathnormal{\pi}}
\wire{An}{P}{3.1}{2.2} \opsymbol{a} \outwire{P}{2.2} \Opsymbol{a}
\end{Diagram}
~~\equiv ~~
\begin{Diagram}{0}{0}
\opbox[5.2]{An}{0,0} \opsymbol{A\mathnormal{[\pi(\negs n\negs)]}}
\outwire{An}{3.1} \Opsymbol{a}
\end{Diagram}
\]
for some permutation $\pi$. That is, a permutation transformation permutes the elements of a maximal set of distinguishable states.  

\subsection{P1}

For the purpose of clarity, it is worth discussing the first postulate at this stage.
\begin{itemize}
\item[{\bf P1}] \emph{Logical Sharpness.}  There is a one-to-one map between pure states and maximal effects such that we get unit probability.
\end{itemize}
This means that for any given pure state there is a unique maximal effect giving unit probability, and that for any given maximal effect there is a unique pure state giving unit probability. In pictures, there is a one-to-one map between pure states and maximal effects:
\[
\begin{Diagram}{0}{-0.2}
\opbox[3]{Us}{0,0} \opsymbol{U}
\outwire{Us}{2} \Opsymbol{a}
\end{Diagram}
~~~\leftrightarrow~~~
\begin{Diagram}{0}{0.2}
\opbox[3]{Us}{0,0} \opsymbol{U}
\inwire{Us}{2} \Opsymbol{a}
\end{Diagram}
~~~~~~~\text{such that}~~~~~~~
\text{Prob}\left(~
\begin{Diagram}{0}{-0.6}
\opbox[3]{Us}{0,0} \opsymbol{U}
\opbox[3]{Ue}{0,4} \opsymbol{U}
\wire{Us}{Ue}{2}{2} \opsymbol{a}
\end{Diagram}
~\right)
=1
\]
Interestingly, causality follows from this postulate.  This is the property that choices in the future do not influence probabilities the past.   The causality property was introduced by CDP as corresponding to the existence of a unique deterministic effect \cite{chiribella2010probabilistic} and used as a postulate in their reconstruction \cite{chiribella2010informational}.

\subsection{Informational faces and non-flat sets of states}

An {\bf informational face}, $S$, is the full set of states having support only on some subset, $O_S$, of the outcomes of some maximal measurement, $\{\mathsf{B_{a_1}}[m] \}$.
Basically, these are sets of states constrained to have a certain information carrying capacity. The states in $\bar{S}$ have support on the complement subset of outcomes, $\bar{O}_S$, for the same maximal measurement.  In convex geometry a face is given by the intersection of the convex set in question and a supporting hyperplane.  A supporting hyperplane is one which has no elements of the convex set on one side.   The supporting hyperplane defining $S$ is given by the equation
\begin{equation}
(\sum_{m\in \bar{O}_S} B_{a_1}[m]) A^{a_1} = 0
\end{equation}
Faces are, themselves, convex sets.  In quantum theory all faces are, in fact, informational faces by virtue of the \emph{spectrality} property (any state can be written as a convex combination of states in a maximal distinguishable set).  However, this need not be the case and we do not assume this here.

A set of states is {\bf non-flat} if it is a spanning subset of some informational face.  It could be an over-complete spanning subset and consequently the informational face is, itself, non-flat.   If {\bf P1} holds then we can think of a non-flat set of states as a kind of generalization of the notion of a pure state.  In fact it follows from {\bf P1} that any single member non-flat set of states consists of a state proportional to a pure state.  Thus, we can think of a set containing a single pure state as being the simplest type of non-flat set in a hierarchy of bigger and bigger non-flat sets.

We need the following notion to understand {\bf P5}.
\begin{quote}
A transformation is said to be {\bf non-flattening} if, for any non-flat set of states we send in, we get a non-flat set of states out.
\end{quote}
It follows from {\bf P1} that all non-flattening transformations are also non-mixing.  Interestingly, in quantum theory the converse is true also: all non-mixing transformations are non-flattening.

\subsection{Filters}

A filter, $\mathsf F$, is defined with respect to a given informational face, $S$.
\begin{quote} {\bf A filter} is a transformation that
\begin{itemize}
\item passes unchanged states in $S$
\item blocks states in $\bar{S}$
\end{itemize}
\end{quote}
For a filter defined with respect to an informational face $S$ given by maximal measurement $\{ \mathsf{B}[m]\}$ and outcome set $O_\mathsf{S}$ we have
\[
\begin{Diagram}{0}{-0.8}
\Opbox[3.4]{A}{0,0} \Opbox[3.4]{F}{0,4}
\wire{A}{F}{2.2}{2.2} \opsymbol{a}
\outwire{F}{2.2} \Opsymbol{a}
\end{Diagram}
~\equiv~
\begin{Diagram}{0}{0}
\Opbox[3.4]{A}{0,0}
\outwire{A}{2.2} \Opsymbol{a}
\end{Diagram}
~~\text{if}~~\text{Prob}\left(
\begin{Diagram}{0}{-0.6}
\Opbox[3.4]{A}{0,0} \opbox[3.4]{Bm}{0,4} \opsymbol{B\mathnormal{[\! m\! ]}}
\wire{A}{Bm}{2.2}{2.2} \opsymbol{a}
\end{Diagram}
\right) =0 ~ \text{for}~m\in \bar{O}_\mathsf{S}
\]

\[
\begin{Diagram}{0}{-0.8}
\Opbox[3.4]{A}{0,0} \Opbox[3.4]{F}{0,4}
\wire{A}{F}{2.2}{2.2} \opsymbol{a}
\outwire{F}{2.2} \Opsymbol{a}
\end{Diagram}
~\equiv~
\begin{Diagram}{0}{0}
\opbox[3.4]{A}{0,0} \opsymbol{0}
\outwire{A}{2.2} \Opsymbol{a}
\end{Diagram}
~~\text{if}~~\text{Prob}\left(
\begin{Diagram}{0}{-0.6}
\Opbox[3.4]{A}{0,0} \opbox[3.4]{Bm}{0,4} \opsymbol{B\mathnormal{[\! m\! ]}}
\wire{A}{Bm}{2.2}{2.2} \opsymbol{a}
\end{Diagram}
\right) =0 ~ \text{for}~m\in {O}_\mathsf{S}
\]
Here
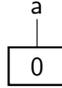
\[
\begin{Diagram}{0}{0}
\opbox[3.4]{A}{0,0} \opsymbol{0}
\outwire{A}{2.2} \Opsymbol{a}
\end{Diagram}
\]
is the preparation corresponding to the null state.  The null state is the state that gives probability zero for any circuit it is part of.  The components of the null state are, therefore, all equal to zero.

\section{THE POSTULATES}

 {Classical probability theory} and  {quantum theory} are only two theories consistent with the following postulates.
\begin{itemize}
\item[{\bf P1}] \emph{Logical sharpness.} There is a one-to-one map between pure states and maximal effects such that we get unit probability.
\item[{\bf P2}] \emph{Information locality.} A maximal measurement on a composite system is effected if we perform maximal measurements on each of the components.  Equivalently $N_\mathsf{ab}=N_\mathsf{a} N_\mathsf{b}$.
\item[{\bf P3}] \emph{Tomographic locality.} The state of a composite system can be determined from the statistics collected by making measurements on the components. Equivalently $K_\mathsf{ab}=K_\mathsf{a} K_\mathsf{b}$.
\item[{\bf P4$'$}] \emph{Permutability.} There exists a reversible transformation on any system effecting any given permutation of any given maximal set of distinguishable states for that system.
\item[{\bf P5}] \emph{Sturdiness.} Filters non-flattening.
\end{itemize}

\subsection{Ruling out the classical case}

To single out quantum theory it suffices to add  \emph{anything} that is inconsistent with classical probability and consistent with quantum theory.  The key property of non-classical theories is that $K_\mathsf{a} > N_\mathsf{a}$ for non-trivial systems (i.e.\ systems having $N_\mathsf{a}>1$).  One way to ensure this is to replace {\bf P4$'$} with
\begin{itemize}
\item[{\bf P4}] \emph{ Compound permutability.} There exists a  {compound} reversible transformation on any system effecting any given permutation of any given maximal set of distinguishable states for that system.
\end{itemize}
A compound transformation is one that can be made from two sequential transformations (neither equal to the identity).  The advantage of this is that it requires only adding a single word (the word \lq\lq compound") to one of the existing postulates.  However, as we just mentioned, we could add any property inconsistent with classical probability theory so long as it is consistent with quantum theory.  For example, we could simply demand that there are more pure states than there are states in any maximal distinguishable set of states for non-trivial systems.

\subsection{P2}

Our second postulate is the following.
\begin{itemize}
\item[{\bf P2}] \emph{Information locality.} A maximal measurement on a composite system is effected if we perform maximal measurements on each of the components.
\end{itemize}
This means the set of results (with $m=1$ to $N_\mathsf{a}$ and $n=1$ to $N_\mathsf{b}$)
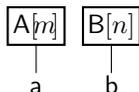
\[
\begin{Diagram}{0}{0}
\opbox[3.4]{Am}{0,0} \opsymbol{A\mathnormal{[\! m \!]}} \inwire{Am}{2.2} \Opsymbol{a}
\opbox[3.4]{Bn}{4,0} \opsymbol{B\mathnormal{[n]}} \inwire{Bn}{2.2} \Opsymbol{b}
\end{Diagram}
\]
is a maximal measurement.

If $N_\mathsf{a}$ is the maximum number of distinguishable states then {\bf P2} is equivalent to statement that
\[ N_\mathsf{ab}=N_\mathsf{a} N_\mathsf{b}  \]
This is very natural.  For example, if we have a die ($N_\mathsf{a}=6$) and a coin ($N_\mathsf{b}=2$) then we have $N_\mathsf{ab}=12$.
We call this \lq\lq information locality" since the total information capacity is given by adding together the local information capacities:
\[ \log N_\mathsf{ab} = \log N_\mathsf{a} + \log N_\mathsf{b}    \]
This postulate is looks innocent but it is actually very powerful. Certainly we can imagine situations in which this postulate is not true (see \cite{hardy2010limited}).

\subsection{P3}

Our third postulate is used in many of the recent reconstructions of quantum theory.  It can be stated in the following way.
\begin{itemize}
\item[{\bf P3}] \emph{Tomographic locality.} The state of a composite system can be determined from the statistics collected by making measurements on the components.
\end{itemize}
Pictorially this means we can determine the state associated with the preparation, $\mathsf{A^{a_1b_2}}$, by determining the probabilities for circuits of the form
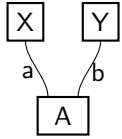
\[
\begin{Diagram}{0}{0}
\Opbox{A}{0,0} \Opbox[2]{X}{-2,5} \Opbox[2]{Y}{2,5}
\wire{A}{X}{1}{1.5} \opsymbol{a} \wire{A}{Y}{3}{1.5} \otherside \opsymbol{b}
\end{Diagram}
\]
It follows from this that we can write the state as $A^{a_1b_2}$ where this is a list of joint probabilities determined by putting separate fiducial results on system $\mathsf a_1$ and $\mathsf b_2$.  In fact, more generally, it follows from tomographic locality that we can represent an arbitrary operation such as $\mathsf{B_{a_1b_2c_3}^{d_4e_5}}$ by a tensor $B_{a_1b_2c_3}^{d_4e_5}$.  Actually, this fact is an equivalent statement of tomographic locality.  In words the equivalent statement is that an arbitrary operation can be fully characterized by local process tomography.  Then the probability for a circuit is given the scalar obtained by contracting over indices where there are wires in the circuit. For example,
\begin{equation}
\text{Prob}(\mathsf{A^{a_1b_2c_3f_6} B_{a_1b_2c_3}^{d_4e_5} C_{d_4} D_{e_5f_6}}) = A^{a_1b_2c_3f_6} B_{a_1b_2c_3}^{d_4e_5} C_{d_4} D_{e_5f_6}
\end{equation}
In \cite{hardy2010formalism, hardy2011reformulating} a tensor such as $B_{a_1b_2c_3}^{d_4e_5}$ correspond to putting a more general object called a duotensor into standard form. Duotensors play an important role in the full reconstruction. However, we will not discuss them further here.  


Another equivalent statement of tomographic locality is that
\[ K_\mathsf{ab} = K_\mathsf{a}K_\mathsf{b}  \]
(where $K_\mathsf{a}$ is number of  probabilities required to specify state).   Hence we see that information locality and tomographic locality are very similar postulates (they were grouped together in \cite{hardy2001quantum}).

There exist other equivalent statements of the tomographic locality assumption (see \cite{hardy2011reformulating} for some of them).

\subsection{P4$'$}

The forth postulate concerns the ability to permute the states in a maximal set of distinguishable states by means of a reversible transformation.
\begin{itemize}
\item[P4$'$] \emph{ Permutability.} There exists a reversible transformation on any system effecting any given permutation of any given maximal set of distinguishable states for that system.
\end{itemize}
In pictures we can say that exists reversible $\mathsf{P}_{\pi}$
\vskip -8mm
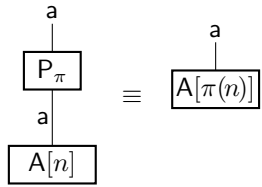
\[
\begin{Diagram}{0}{-1}
\opbox[5.4]{An}{0,0} \opsymbol{A\mathnormal{[n]}}
\opbox[3.4]{P}{0,5}   \opsymbol{P_\mathnormal{\pi}}
\wire{An}{P}{3.2}{2.2} \opsymbol{a} \outwire{P}{2.2} \Opsymbol{a}
\end{Diagram}
~~\equiv ~~
\begin{Diagram}{0}{0}
\opbox[5.4]{An}{0,0} \opsymbol{A\mathnormal{[\pi(\negs n\negs)]}}
\outwire{An}{3.2} \Opsymbol{a}
\end{Diagram}
\]
for any maximal set of distinguishable states, $A^{a_1}[n]$ and permutation, $\pi$.  Reversibility means that this transformation is reversible when applied to any state (not just the members of the maximal set of distinguishable states).  

This postulate implies we can perform lossless arbitrary translation of a message encoded with respect to any alphabet to one encoded with respect to any permutation of this same alphabet.

\subsection{P5}

The last postulate concerns filters.
\begin{itemize}
\item[P5] \emph{Sturdiness.} Filters are non-flattening.
\end{itemize}
Recall that a set of states is said to be {\bf non-flat} if it is a spanning subset some informational face.  One way to think of this property is that sets of states resist being squashed (hence the name \lq\lq sturdiness").  Quantum states are not as sensitive as we might have imagined.  A filter is a pretty dramatic transformation. However, according to this postulates, sets of states remain as intact as they can under the circumstances.

\section{Outline of reconstruction}

The full reconstruction, while only using elementary mathematics, is rather lengthy.  Here we will only give an outline of some of the main steps.

First, using {\bf P1}, {\bf P2}, {\bf P3}, and {\bf P4$'$} we
\begin{itemize}
\item show that there exists a reversible transformation between any pair of pure states,
\item construct arbitrary filters,
\item show there exist types with $N=1, 2, 3, \dots$,
\item show that systems having same $N$ are equivalent,
\item show that $K_\mathsf{a}=N^r_\mathsf{a}$ where $r=1, 2, 3, \dots$ (the Wootters hierarchy \cite{wootters1986quantum, wootters1990local}.
\end{itemize}
Using {\bf P5} as well we 
\begin{itemize}
\item show that gebits (generalized bits, i.e.\ systems having $N_\mathsf{a}=2$) correspond to hyperspheres,
\item show that all points on the hypersphere correspond to pure states,
\item show how to do teleportation,
\item prove that $K_\mathsf{a}=N_\mathsf{a}$ or $K_\mathsf{a}=N^2_\mathsf{a}$.
\end{itemize}
This gives us the bit or the qubit.  Interestingly, getting the qubit is the most difficult part of this reconstruction as well as many others.  Having got the qubit we get the appropriate constraints on quantum theory in general (not just the $N_\mathsf{a}=2$ case) by showing that a certain \lq\lq magic operation" can implement any complete set of superoperators (superoperators corresponding to a set of operations associated with a given apparatus with disjoint outcome sets where the union of these outcome sets is the full set of outcomes).  To complete this last step and get quantum theory in general we employ the duotensor framework \cite{hardy2010formalism} and the operator tensor formulation of quantum theory \cite{hardy2011reformulating, hardy2012operator}.  We will provide an outline of how some aspects of the reconstruction work in the following subsections.  For the full details of the proofs (which are mostly omitted here) the reader is referred to \cite{hardy2011reformulating}.

\subsection{Reversible transformation between pure states}

Let
\[ \{ \mathsf{U^{a_1}}[n]: n=1~\text{to}~N_\mathsf{a} \}  ~~~~~~\text{and}~~~~~ \{ \mathsf{V^{a_1}}[n]: n=1~\text{to}~N_\mathsf{a} \} \]
be maximal sets of distinguishable preparations for $\mathsf a$.  Let
\[ \{ \mathsf{W^{b_2}}[m]: m=1~\text{to}~N_\mathsf{b} \} \]
be a maximal set of distinguishable preparations for $\mathsf b$.  We will denote the maximal measurement that distinguishes these maximal sets of distinguishable states by $\{ \mathsf{U_{a_1}}[n] \}$, $\{ \mathsf{V_{a_1}}[n] \}$, and $\{ \mathsf{W_{b_2}}[m] \}$.  

It follows from {\bf P2} that
\[ \{ \mathsf{U^{a_1}}[n]\mathsf{W^{b_2}}[m]: nm=11, 12, \dots N_\mathsf{a}N_\mathsf{b} \} \]
is a maximal set for $\mathsf{ab}$.  Similarly,
\[ \{ \mathsf{V^{a_1}}[n]\mathsf{W^{b_2}}[m]: nm=11, 12, \dots N_\mathsf{a}N_\mathsf{b} \} \]
is another maximal set for $\mathsf{ab}$.  

Let $\mathsf{P}$ be the reversible transformation that permutes $\mathsf{U^{a_1}}[n]\mathsf{W^{b_2}}[m]$ according to
\[
\pi_\mathsf{P} =  (nm \leftrightarrow mn )
\]
Let $\mathsf Q$ be the reversible transformation that permutes $\mathsf{V^{a_1}}[n]\mathsf{W^{b_2}}[m]$ according to
\[
\pi_\mathsf{Q} = (nm \leftrightarrow mn )
\]
Note we choose $\mathsf b$ such that $N_\mathsf{b}= N_\mathsf{a}$.
Then a little thought shows that
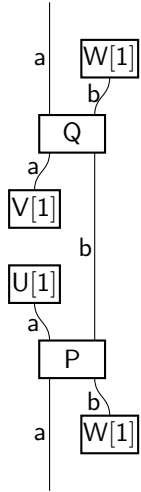
\[
\begin{diagram}
\Opbox[4]{P}{0,0}
\opbox[4]{Q}{0,12} \opsymbol{Q}
\thispoint{IN}{-1.2, -7}   \wire{IN}{P}{1}{1} \opsymbol{a}
\opbox[3.4]{B1}{2,-4} \opsymbol{W[1]}
\wire{B1}{P}{2.2}{4} \opsymbol{b}
\opbox[3]{C1}{-2,4} \opsymbol{U[1]}
\wire{P}{C1}{1}{2} \opsymbol{a}
\wire{P}{Q}{4}{4} \opsymbol{b}
\opbox[3]{A1}{-2,8} \opsymbol{V[1]}
\wire{A1}{Q}{2}{1}  \opsymbol{a}
\opbox[3.4]{D1}{2,16} \opsymbol{W[1]}
\wire{Q}{D1}{4}{2.2} \opsymbol{b}
\thispoint{OUT}{-1.2, 19} \wire{Q}{OUT}{1}{1} \opsymbol{a}
\end{diagram}
\]
does the job. We can prove that this is a reversible transformation and it clearly takes $\mathsf{U^{a_1}}[1]$ to $\mathsf{V^{a_3}}[1]$.  In fact it actually does a bit more. It takes $\mathsf{U^{a_1}}[n]$ to $\mathsf{V^{a_3}}[n]$ for $n=1$ to $N_\mathsf{a}$.

\subsection{Arbitrary filters are possible}

It can be shown that the transformation
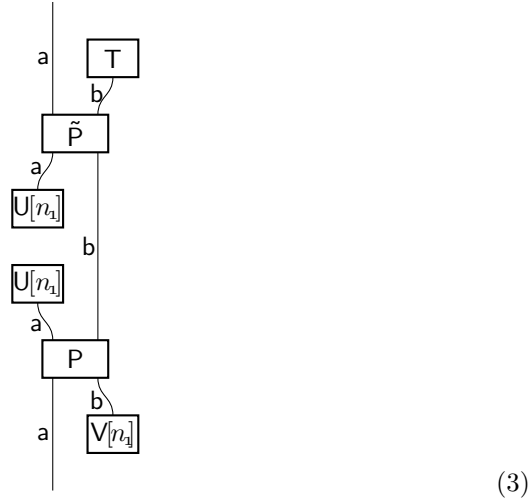
\begin{equation}\label{filter}
\begin{diagram}
\Opbox[4]{P}{0,0}
\opbox[4]{Q}{0,12} \opsymbol{\tilde P}
\thispoint{IN}{-1.2, -7}   \wire{IN}{P}{1}{1} \opsymbol{a}
\opbox[3]{B1}{2,-4} \opsymbol{V\negs[\negs \mathnormal{n_{\!1}\!}\negs]}
\wire{B1}{P}{2}{4} \opsymbol{b}
\opbox[3]{C1}{-2,4} \opsymbol{U\negs[\mathnormal{n_{\!1}\!}]}
\wire{P}{C1}{1}{2} \opsymbol{a}
\wire{P}{Q}{4}{4} \opsymbol{b}
\opbox[3]{A1}{-2,8} \opsymbol{U\negs[\mathnormal{n_{\!1}\!}]}
\wire{A1}{Q}{2}{1}  \opsymbol{a}
\opbox[3]{D1}{2,16} \opsymbol{T}
\wire{Q}{D1}{4}{2} \opsymbol{b}
\thispoint{OUT}{-1.2, 19} \wire{Q}{OUT}{1}{1} \opsymbol{a}
\end{diagram}
\end{equation}
effects an arbitrary filter where $N_\mathsf{a}=N_\mathsf{b}$ and $n_1$ is any integer chosen from $O_S$.  Here the transformation $\mathsf{P}$ is a permutation transformation with permutation
\begin{equation}
\pi = \left( \begin{array}{l} nm \leftrightarrow mn  ~~\text{if} ~~ n~\text{and}~m\in O(S) \\
                              nm \leftrightarrow nm ~~ \text{otherwise}                  \end{array} \right)
\end{equation}
and $\mathsf{\tilde P}$ is the transformation that reverses $\mathsf{P}$.  The result $\mathsf{T_{b}}$ is a deterministic result (its outcome set is equal to the set of all outcomes).

\subsection{Systems with same are $N$ equivalent}

These substitutions prove equivalence when $N_\mathsf{a}=N_\mathsf{b}$.
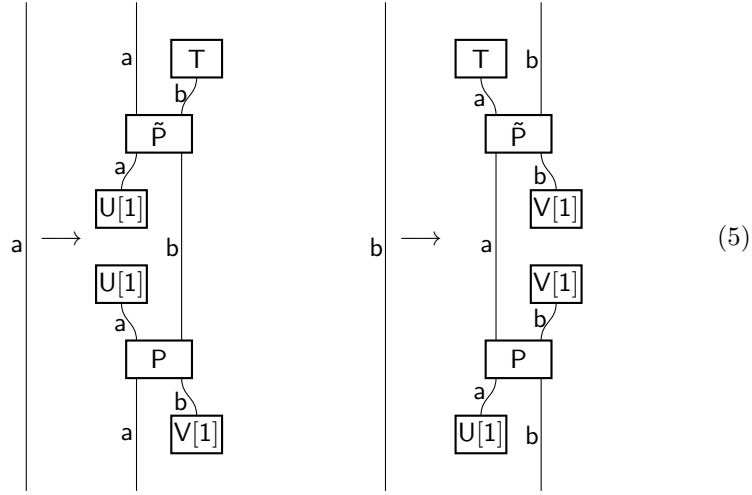
\begin{equation}\label{absubs}
\begin{Diagram}{0}{-1.7}
\thispoint{IN}{0,-7}
\thispoint{OUT}{0,19}
\wire{IN}{OUT}{1}{1} \opsymbol{a}
\end{Diagram}
\longrightarrow
\begin{Diagram}{0}{-1.7}
\Opbox[4]{P}{0,0}
\opbox[4]{Q}{0,12} \opsymbol{\tilde P}
\thispoint{IN}{-1.2, -7}   \wire{IN}{P}{1}{1} \opsymbol{a}
\opbox[3]{B1}{2,-4} \opsymbol{V[1]}
\wire{B1}{P}{2}{4} \opsymbol{b}
\opbox[3]{C1}{-2,4} \opsymbol{U[1]}
\wire{P}{C1}{1}{2} \opsymbol{a}
\wire{P}{Q}{4}{4} \opsymbol{b}
\opbox[3]{A1}{-2,8} \opsymbol{U[1]}
\wire{A1}{Q}{2}{1}  \opsymbol{a}
\opbox[3]{D1}{2,16} \opsymbol{T}
\wire{Q}{D1}{4}{2} \opsymbol{b}
\thispoint{OUT}{-1.2, 19} \wire{Q}{OUT}{1}{1} \opsymbol{a}
\end{Diagram}
~~~~~~~~~~~~~~~
\begin{Diagram}{0}{-1.7}
\thispoint{IN}{0,-7}
\thispoint{OUT}{0,19}
\wire{IN}{OUT}{1}{1} \opsymbol{b}
\end{Diagram}
\longrightarrow
\begin{Diagram}{0}{-1.7}
\opbox[4]{P}{0,0}  \opsymbol{P}
\opbox[4]{Q}{0,12} \opsymbol{\tilde P}
\thispoint{IN}{1.2, -7}
\wire{IN}{P}{1}{4} \opsymbol{b}
\opbox[3]{B1}{-2,-4} \opsymbol{U[1]}
\wire{B1}{P}{2}{1} \opsymbol{a}
\opbox[3]{C1}{2,4} \opsymbol{V[1]}
\wire{P}{C1}{4}{2} \opsymbol{b}
\wire{P}{Q}{1}{1} \opsymbol{a}
\opbox[3]{A1}{2,8} \opsymbol{V[1]}
\wire{A1}{Q}{2}{4}  \opsymbol{b}
\opbox[3]{D1}{-2,16} \opsymbol{T}
\wire{Q}{D1}{1}{2} \opsymbol{a}
\thispoint{OUT}{1.2, 19} \wire{Q}{OUT}{4}{1} \opsymbol{b}
\end{Diagram}
\end{equation}
where
\begin{equation}
\pi = ( nm \leftrightarrow mn  )
\end{equation}
With these substitutions we can replace any wire of type $\mathsf{a}$ by one of type $\mathsf{b}$ (and vice versa) without changing the probability for the given circuit.

\subsection{Proof that $K_\mathsf{a}=N^r_\mathsf{a}$}

It follows from the first four postulates that
\begin{itemize}
\item $K_\mathsf{a}=K(N_\mathsf{a})$ (since systems having the same $N_\mathsf{a}$ are equivalent).
\item $K(N+1)>K(N)$ (since we can filter systems down).
\item $K(N_\mathsf{a}N_\mathsf{b})= K(N_\mathsf{a}) K( N_\mathsf{b})$ (by {\bf P2}).
\end{itemize}
It can be shown that
\vskip 5mm
\[ K_\mathsf{a}=N_\mathsf{a}^r   ~~~\text{where}~r=1,2,3,\dots  \]
follows (the proof of this uses the decomposition of $N_\mathsf{a}$ and $N_\mathsf{b}$ into prime numbers).  This relationship was first suggested by Wootters \cite{wootters1986quantum, wootters1990local} and hence we term it the Wootters hierarchy.  It was first proven that this relationship follows from the above more basic premises in \cite{hardy2001quantum}

\subsection{All points on hypersphere correspond to pure states}

It can be shown to follow from fact that there exists a reversible group of transformations between pure states that all pure states must live on surface of a hypersphere.  We then need to show that all points on this hypersphere correspond to pure states for the gebit.   The proof of this starts with a getrit (a system having $N_\mathsf{a}=3$).  We prepare a system constrained to an $N_\mathsf{a}=2$ informational face of this getrit (i.e.\ we consider states contrained to a gebit space).  Next we filter on the getrit space but with respect to a maximal measurement that has one maximal result having full support on the afore mentioned gebit and one maximal result having only partial support on this gebit.  What happens is that the states emerging out of this filter are also gebit states but they move closer to one of the poles of the gebit (the one associated with the maximal result having full support).  If we keep filtering like this then an intially non-flat set of states will, by {\bf P5} remain non-flat but will move closer to this pole.  We can produce a spanning set of states that are as close to the pole as we like.  It then follows that within an infintessimal region of the pole there must be points lying in any direction.  Since any state could serve as the pole, this proves that all points on the hypersphere are populated.

\subsection{Getting the qubit and $K_\mathsf{a}=N_\mathsf{a}^2$}

The classical case corresponds to the 1-sphere with just two pure states.   If we are in the non-classical case then we want to prove that this hypersphere must be the 2-sphere corresponding to the qubit of quantum theory.   Consequently, we want to prove that for $N_\mathsf{a}=2$ we have $K_\mathsf{a}=4$ (in the non-classical case). This means that we get the Bloch ball (since one parameter counted in $K_\mathsf{a}$ corresponds to normalization).  This proof is adopted from a beautiful proof due to Chiribella, D'Ariano, and Perinotti \cite{chiribella2010probabilistic, chiribella2010informational} using teleportation.

We start by assuming that we are in the non-classical case.  Consider the gebit preparation
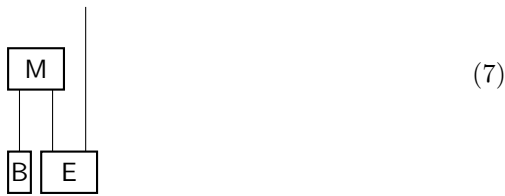
\begin{equation}\label{teleportationone}
\begin{Diagram}[1.1]{0}{0}
\Opbox{M}{0,0}
\Opbox{E}{1.6,-5}
\Opbox[1]{B}{-0.8,-5}
\wire{B}{M}{1}{1}
\wire{E}{M}{1}{3}
\thispoint{out}{2.4,3}
\wire{E}{out}{3}{1}
\end{Diagram}
\end{equation}
where $\mathsf{B^{a_1}}$ is a gebit preparation, $E^{a_2a_3}$ is an entangled pure state in $S_{\{11,22\} }$ and $M_{a_1a_2}$ is a certain maximal entangled effect (these do not exist in the classical case but must exist in the non-classical case).  We can show that the transformation on $B^{a_1}$ is non-flattening using {\bf P5}.  The pure states for preparation ${\mathsf B^{a_1}}$ lie on the surface of a hypersphere.  Under the transformation in (\ref{teleportationone}) this hypersphere is transformed to an hyper-ellipsoid.  Hence we can use the preparation in (\ref{teleportationone}) to prepare a state proportional to any pure state by making an appropriate choice of preparation $\mathsf B$.  The state that is prepared is not necessarily equal to $B^{a_1}$ so we do not necessarily have faithful teleportation.  However, we can use this result to obtain the following result.
\begin{equation}
\begin{Diagram}[1.1]{0}{0}
\Opbox{M}{-1.6,3}    \Opbox[1]{1}{0.8,3}
\Opbox[1]{B}{-2.4,0}     \opbox{Pt}{0,0} \opsymbol{\tilde P_\text{\negs\negs\negs cnot}}
\wire{B}{M}{1}{1}  \wire{Pt}{M}{1}{3} \wire{Pt}{1}{3}{1}
\Opbox[1]{A}{-2.4,-5} \wire{A}{Pt}{1}{1}
\opbox{M1}{1.6,-5} \opsymbol{M}
\wire{M1}{Pt}{1}{3}
\thispoint{out}{2.4,5} \wire{M1}{out}{3}{1}
\end{Diagram}
~~~\equiv~ \frac{1}{8}~
\begin{Diagram}[1.1]{0}{0}
\Opbox[1]{A}{0,0}
\thispoint{out}{0,5}
\wire{A}{out}{1}{1}
\end{Diagram}
\end{equation}
for any state $A^{a_1}$ for the gebit.  Now we do have faithful (probabilistic) teleportation.  We work in a computational basis for the gebit denoted by 0 and 1.  Here $B^{a_2}$ is a special choice of state.   In fact it must be an {\it equatorial state} - a pure state on the equator of the hypersphere between the two poles (these exist only in the non-classical case).  Also, $\mathsf{P}_{\text{cnot}}$ is a reversible permutation transformation effecting the permutation associated with the $CNOT$ gate in the computational basis.

Since we now have faithful (albeit probabilistic) teleportation, we have
\[
\begin{Diagram}[1.1]{0}{0}
\Opbox{M}{-1.6,3}    \Opbox[1]{1}{0.8,3}
\Opbox[1]{B}{-2.4,0}     \opbox{Pt}{0,0} \opsymbol{\tilde P_\text{\negs\negs\negs cnot}}
\wire{B}{M}{1}{1}  \wire{Pt}{M}{1}{3} \wire{Pt}{1}{3}{1}
\thispoint{A}{-0.8,-7} \wire{A}{Pt}{1}{1}
\opbox{M1}{1.6,-5} \opsymbol{M}
\wire{M1}{Pt}{1}{3}
\thispoint{out}{2.4,5} \wire{M1}{out}{3}{1}
\end{Diagram}
~~\equiv~\frac{1}{8}
\begin{Diagram}[1.1]{0}{-1.8}
\thispoint{in}{0,0} \thispoint{out}{2,11} \wire[2]{in}{out}{1}{1}
\end{Diagram}
\]
We can also prove that
\begin{equation}\label{closeloop}
\text{Prob}\left(
\begin{Diagram}[1.1]{0}{0.15}
\Opbox{M}{-1.6,3}    \Opbox[1]{1}{0.8,3}
\Opbox[1]{B}{-2.4,0}     \opbox{Pt}{0,0} \opsymbol{\tilde P_\text{\negs\negs\negs cnot}}
\wire{B}{M}{1}{1}  \wire{Pt}{M}{1}{3} \wire{Pt}{1}{3}{1}
%
%
\opbox{M1}{1.6,-5} \opsymbol{M}
\wire{M1}{Pt}{1}{3}
%
\wire{M1}{Pt}{3}{1}
\end{Diagram}
~\right)
\leq ~ \frac{1}{2}
\end{equation}
using the fact that $B^{a_2}$ is equatorial.   For convenience we put
\begin{equation}
\begin{Diagram}[1.1]{0}{0}
\Opbox{N}{0,0}
\inwire{N}{1}\inwire{N}{3}
\end{Diagram}
~~:=~~
\begin{Diagram}[1.1]{0}{0}
\Opbox{M}{-1.6,3}    \Opbox[1]{1}{0.8,3}
\Opbox[1]{B}{-2.4,0}     \opbox{Pt}{0,0} \opsymbol{\tilde P_\text{\negs\negs\negs cnot}}
\wire{B}{M}{1}{1}  \wire{Pt}{M}{1}{3} \wire{Pt}{1}{3}{1}
\inwire{Pt}{1} \inwire{Pt}{3}
\end{Diagram}
\end{equation}
Then
\begin{equation}
N_{a_1a_2}M^{a_2a_3}= \frac{1}{8} I_{a_1}^{a_3}
\end{equation}
where $I_{a_1}^{a_3}$ is the identity. Hence,
\begin{equation}
N_{a_1a_2}M^{a_2a_1}= \frac{1}{8} I_{a_1}^{a_1}= \frac{1}{8} K_\mathsf{a}
\end{equation}
since the trace of the identity is equal to the dimension of the space on which it acts.  But we also have
\begin{equation}
N_{a_1a_2}M^{a_2a_1} \leq \frac{1}{2}
\end{equation}
It follows that, for a gebit, $K_\mathsf{a}\leq 4$. Hence, $K=N^2$ in general.

\subsection{The magic operation}

The last part of the proof shows how to use the fact that the gebit is equal to the qubit along with the postulates to get quantum theory in general.  The key part of this is showing that the following set of operations (for different outcomes $l$ of a maximal measurement) 
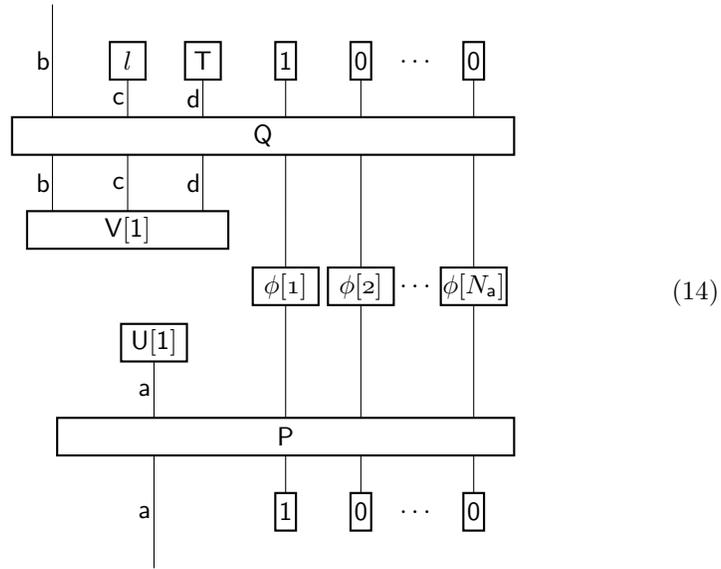
\begin{equation}\label{magic}
\begin{Diagram}{0}{-2}
\opbox[1]{zero1}{0,-4} \opsymbol{1} \opbox[1]{zero2}{4,-4} \opsymbol{0} \opbox[1]{zero3}{10,-4} \opsymbol{0}
\Opbox[30]{P}{0,0}
\opbox[4]{U1}{-7,5} \opsymbol{U[1]}
\opbox[4]{phi1}{0,8} \opsymbol{\mathnormal{\phi[1]}}
\opbox[4]{phi2}{4,8} \opsymbol{\mathnormal{\phi[2]}}
\opbox[4]{phi3}{10,8} \opsymbol{\mathnormal{\phi[N_\mathsf{a}]}}
\opbox[13]{V1}{-8.4,11} \opsymbol{V[1]}
\opbox[2]{aux1t}{-8.4,20} \opsymbol{\mathnormal{l}}
\opbox[2]{aux2t}{-4.4,20} \opsymbol{T}
\Opbox[33]{Q}{-1.2,16}
\opbox[1]{zero1t}{0,20} \opsymbol{1} \opbox[1]{zero2t}{4,20} \opsymbol{0} \opbox[1]{zero3t}{10,20} \opsymbol{0}
\thispoint{in}{-7,-7} \thispoint{out}{-12.4,23}
\wire{in}{P}{1}{6.75} \opsymbol{a}
\wire{Q}{out}{3}{1}\opsymbol{b}
\wire{P}{U1}{6.75}{2.5} \opsymbol{a}
\wire{V1}{Q}{2}{3} \opsymbol{b}
\wire{V1}{Q}{7}{8} \opsymbol{c}
\wire{V1}{Q}{12}{13} \opsymbol{d}
\wire{Q}{aux1t}{8}{1.5} \opsymbol{c}
\wire{Q}{aux2t}{13}{1.5} \opsymbol{d}
\wire{zero1}{P}{1}{15.5} \wire{zero2}{P}{1}{20.5} \wire{zero3}{P}{1}{28}
\wire{Q}{zero1t}{18.5}{1} \wire{Q}{zero2t}{23.5}{1} \wire{Q}{zero3t}{31}{1}
\wire{P}{phi1}{15.5}{2.5} \wire{P}{phi2}{20.5}{2.5} \wire{P}{phi3}{28}{2.5}
\wire{phi1}{Q}{2.5}{18.5} \wire{phi2}{Q}{2.5}{23.5} \wire{phi3}{Q}{2.5}{31}
\placelatex{7,-4}{\dots}  \placelatex{7,8}{\dots}\placelatex{7,20}{\dots}
\end{Diagram}
\end{equation}
can generate any complete set of operations in quantum theory.  Here $\mathsf P$ and $\mathsf Q$ are appropriately chosen reversible permutation transformations, $\{\phi[n]: n=1~\text{to}~N_\mathsf{a}\}$ are appropriately chosen phases, $\mathsf c$ and $\mathsf d$ are ancillary systems having appropriate $N_\mathsf{c}$ and $N_\mathsf{d}$, and $\mathsf{V^{b_2c_3d_4}}[1]$ is an appropriately chosen preparation (for a pure state).  The unlabeled wires represent qubits. $\mathsf T$ is the deterministic effect.  This proof employs the duotensor and operator tensor frameworks and the reader is referred to \cite{hardy2011reformulating} for details.

\section{Conclusions}

I have provided a set of operational postulates from which quantum theory can be reconstructed. This does not require a simplicity assumption as did my earlier work \cite{hardy2001quantum} from over a decade ago.   This is one of a number of recent reconstructions \cite{dakic2009quantum, chiribella2010informational, masanes2010derivation, masanes2012digital, zaopo2012information} along similar lines which use the assumption of tomographic locality (like \cite{hardy2001quantum}) and do not need a simplicity assumption.   There are strong connections between these different approaches and many of the proof techniques are similar.  What appears as a postulate in one approach appears as a low level theorem in another and vice versa.  One might think of a set of postulates as being a little akin to a choice of coordinate system used to represent some shape. If we find a good coordinate system then the shape appears simple.  The fact that there are a number of good postulate sets that are fairly simply related to each other is similar to the fact that there are often a number of good choices of simply related coordinate system for viewing a shape.   While one may have preferences for one or the other set of postulates, there is not really much to distinguish them.

However, I am left with the sense that some much deeper insights are still left to be had.  One reason for this sense is that in the operator tensor formulation \cite{hardy2011reformulating, hardy2012operator} of quantum theory (and, similarly, in the quantum combs approach \cite{chiribella2009theoretical}) preparations, transformations, and results are all treated on a fairly equal footing. However, this is not true of the postulate set presented here or the others I have mentioned.  Surely the postulates should also treat all kinds of operations on an  equal footing (so far as this is possible).  Further motivation for this comes from quantum gravity.  We do not yet have a theory of quantum gravity. However, when we do, it seems likely that we will have to contend with indefinite causal structure.  We will not be able to say whether some particular interval is space-like, time-like, or null, but rather can expect to have something like a quantum superposition of these different cases.  Then we cannot be sure that some ports on an operation are inputs and others are outputs (since these notions assume definite causal structure).  In this case we cannot distinguish preparations, transformations, and results.  Quantum theory might reasonably be expected to be obtained as a limit of quantum gravity (in the limit as we have definite causal structure).  In this limit distinct notions of preparations, transformations, and results might emerge. However, fundamentally (before the limit is taken), they are not distinct.  It would, then, be great if a set of postulates for quantum theory treated them more-or-less on an equal footing. Some of these postulates may then go over to a theory of quantum gravity. In \cite{hardy2005probability, hardy2007towards, hardy2009quantum3} I developed the \lq\lq causaloid framework" for general probabilistic theories that can accommodate indefinite causal structure.  One can put quantum theory into this framework and it does, indeed, have the feature that preparations, transformations, and results are then treated on a (more-or-less) equal footing (indeed, the causaloid formulation of quantum theory is the origin of the operator tensor formulation).

I think that the real test of this research program will be the progress that is made towards a theory of quantum gravity using these newly developed techniques.  It is in constructing new physical theories that we can really test whether we are on the right path since then we have to make new predictions and account for new experimental data.  Furthermore, the more fundamental our physical theory, the more natural we can expect our postulates to be.

\section*{Acknowledgements}

First and foremost I am grateful to Christopher Fuchs for getting me thinking about these questions in the first place.  He is responsible for initiating the present wave of reconstruction work.  I am also grateful to Guilio Chiribella, Mauro D'Ariano, and Paulo Perinotti both for discussions on these matters and for writing a number of inspirational papers.  Over the years I have benefitted by discussions with many people on the subject of reconstructing quantum theory to whom I am grateful.

Research at Perimeter Institute is supported by the Government 
of Canada through Industry Canada and by the Province of Ontario through the Ministry
of Economic Development and Innovation.

\bibliography{quantum}
\bibliographystyle{plain}






\end{document}